\documentclass[aps,two column, prd,reprint,superscriptaddress,amssymb]{revtex4-1}

\usepackage{amsmath}

\usepackage{graphicx}
\usepackage{floatrow}
\usepackage[leftcaption]{sidecap}
\sidecaptionvpos{figure}{right}
\usepackage{amsmath,amsfonts,amsbsy,amssymb,array,accents,dsfont}
\usepackage{enumerate,array,latexsym,graphicx,mathrsfs,verbatim,psfrag}
\usepackage{bm} %math bold symbols
\usepackage[normalem]{ulem}%strikethrough
\usepackage{multirow}
\usepackage{cancel}
\usepackage{chngpage} 

\usepackage{booktabs} %some alignment stuff
\usepackage[usenames]{color}
\usepackage[utf8]{inputenc}
\usepackage[english]{babel}
\usepackage{fancybox}
\usepackage[dvipsnames]{xcolor}

\usepackage{tikz}
\usepackage{pgfplots}
\usetikzlibrary{decorations.text,intersections, pgfplots.fillbetween}
\usetikzlibrary{math}
 \usetikzlibrary{snakes,3d,shapes.geometric,shadows.blur}
\usepackage{tikz-3dplot}

\usetikzlibrary{positioning}
\usetikzlibrary{automata}
\usetikzlibrary{arrows}
\usetikzlibrary{calc}
\usetikzlibrary{decorations.markings}
\usetikzlibrary{decorations.pathreplacing}
\usetikzlibrary{intersections}
\usetikzlibrary{positioning}
\usetikzlibrary{topaths}
\usetikzlibrary{shapes.geometric}
\usetikzlibrary{shapes.misc}
\tikzset{cf-group/.style = {
    shape = rounded rectangle, minimum size=1.0cm,
    rotate=90,
    rounded rectangle right arc = none,
    draw}}
\tikzset{cross/.style={path picture={ 
  \draw[black]
(path picture bounding box.south east) -- (path picture bounding box.north west) (path picture bounding box.south west) -- (path picture bounding box.north east);
}}}

\tikzset{unode/.style=
{black, circle,draw,thick,fill=black!100 ,minimum size=1mm}}

\tikzset{sunode/.style=
{black, circle,draw,thick,fill=yellow!100 ,minimum size=1mm}}

\tikzset{fnode/.style=
{black, rectangle,draw,thick,minimum size=1mm}}

\tikzset{afnode/.style=
{blue,rectangle,draw,thick,minimum size=1mm}}

\tikzset{rnode/.style={black, circle,draw,thick,fill=gray!100 ,minimum size=5mm}}

\usepackage{sidecap}

\sidecaptionvpos{figure}{c}

\usepackage{mciteplus}

\usepackage[colorlinks=true,      linkcolor=blue,      urlcolor=blue,      
            filecolor=blue,      citecolor=blue,       pdfstartview=FitH,     
						pdfpagemode=UseNone,      bookmarksopen=true]{hyperref}  % LINKS
\usepackage[all]{hypcap}     %should be loaded AFTER hyperref, which should otherwise be loaded last.

\newcommand{\be}{\begin{equation}}
\newcommand{\ee}{\end{equation}}
\newcommand{\ba}{\begin{array}}
\newcommand{\ea}{\end{array}} 
\newcommand{\bi}{\begin{itemize}}
\newcommand{\ei}{\end{itemize}}
\def\vec#1{\bm{#1}}
\def\bea#1\eea{\allowdisplaybreaMs \begin{align}#1\end{align}}
 \newcommand{\ben}{\begin{enumerate}}
\newcommand{\een}{\end{enumerate}}
\newcommand{\bean}{\begin{eqnarray*}}
\newcommand{\eean}{\end{eqnarray*}}
\newcommand{\eref}[1]{(\ref{#1})}

\newcommand{\nn}{\nonumber}

\newcommand{\CS}{{\cal S}}

\newcommand{\CT}{{\cal T}}

\newcommand{\CO}{{\cal O}}

\newcommand{\CN}{{\cal N}}

\newcommand{\frsu}{\mathfrak{su}}
\newcommand{\fru}{\mathfrak{u}}

\newcommand{\frg}{\mathfrak{g}}

\newcommand{\wt}{\widetilde}

\newcommand{\Figref}[1]{Figure~\ref{#1}}
\newcommand{\figref}[1]{Fig.~\ref{#1}}
\renewcommand{\eqref}[1]{(\ref{#1})}

\usepackage{titlesec}
\titlespacing*{\section}
{0pt}{0.5cm}{0.2cm}
\titlespacing*{\subsection}
{0pt}{0.25cm}{0.1cm}

\begin{document}

\title{Three dimensional Mirrors from IR N-ality}

\author{Anindya Dey}
\email{anindya.hepth@gmail.com}
\affiliation{Department of Physics and Astronomy, Johns Hopkins University, 3400 North Charles Street, Baltimore, MD 21218, USA}

\begin{abstract}

We discuss an explicit field theory construction of three dimensional mirrors for a large sub-class of quiver gauge theories involving 
unitary and special unitary gauge nodes with matter in fundamental and bifundamental representations. For this sub-class of theories, one 
can deploy a sequence of IR dualities and certain Abelian field theory operations to reduce the aforementioned problem to a 
problem of finding the 3d mirror of a simpler unitary quiver gauge theory. We illustrate the construction with a linear quiver 
consisting of unitary and special unitary gauge nodes. 
\end{abstract}

\maketitle

%\section{Introduction} \label{Intro}

\noindent \textit{Introduction.}  3d $\CN=4$ theories have a rich landscape of IR dualities, the most well-known of them being mirror symmetry \cite{Intriligator:1996ex, deBoer:1996mp}. This duality acts on a pair of 
quiver gauge theories, which are distinct in the UV in terms of the gauge groups and the matter content but flow to the same SCFT in the IR, by exchanging the Coulomb branch (CB) of one theory with the Higgs branch (HB) of the other 
(and vice-versa) in the deep IR. Mirror symmetry is the supersymmetric parent of a very large class of non-supersymmetric boson/fermion-type 
dualities which arise from the former by a soft supersymmetry-breaking mechanism \cite{Kachru:2016rui}. 
Historically, the construction of mirror duals has relied heavily on the realization of these theories in terms of brane systems in String Theory and 
deploying String dualities. The Hanany-Witten \cite{Hanany:1996ie} construction in the Type IIB String Theory is the earliest avatar of this procedure. 
For a large class of quiver gauge theories, 3d mirrors have also been constructed via the magnetic quiver approach which makes use of 
five-dimensional brane webs \cite{Bourget:2019rtl, Bourget:2021jwo}.

Recently, there have been renewed efforts to probe 3d mirror symmetry from an exclusively field theory perspective \cite{Dey:2020hfe, Dey:2021jbf, Hwang:2020wpd, Hwang:2021ulb}. 
In particular, \cite{Dey:2020hfe} focussed on constructing mirror duals for quiver gauge theories with unitary gauge nodes given by a 
generic graph, provided the theories are good in the Gaiotto-Witten sense \cite{Gaiotto:2008ak}. 
While these theories generically do not have a known string theory realization, one can construct the 3d mirrors of a 
large subclass in a systematic fashion starting from the well-understood duality of linear ($A$-type) quivers, using a generalized version of 
the $S$-operation \cite{Witten:2003ya} on a 3d CFT. \\

\begin{figure}[htbp]
\begin{center}
\scalebox{0.55}{\begin{tikzpicture}
\node[] (100) at (-3,0) {};
\node[] (1) at (-1,0) {};
\node[unode] (2) at (0,0) {};
\node[text width=.2cm](31) at (0.1,-0.5){$N_1$};
\node[sunode] (3) at (2,0) {};
\node[text width=.2cm](32) at (2.1,-0.5){$N_2$};
\node[] (4) at (3,0) {};
\node[] (5) at (4,0) {};
\node[unode] (6) at (5,0) {};
\node[text width=.2cm](33) at (5.1,-0.5){$N_{\alpha}$};
\node[unode] (7) at (7,0) {};
\node[text width=.2cm](34) at (7.1,-0.5){$N_{\alpha+1}$};
\node[sunode] (8) at (9,0) {};
\node[text width=.2cm] (40) at (9.1,-0.5) {$N_{\alpha+2}$};
\node[fnode] (9) at (9,-2) {};
\node[text width=.2cm] (40) at (9.5,-2) {$M_{\alpha+2}$};
\node[] (10) at (10,0) {};
\node[] (11) at (12,0) {};
\node[fnode] (20) at (0,-2) {};
\node[fnode] (21) at (2,-2) {};
\node[fnode] (22) at (5,-2) {};
\node[fnode] (27) at (7,-2) {};
\node[text width=.2cm](23) at (0.5,-2){$M_1$};
\node[text width=.2cm](24) at (2.5,-2){$M_2$};
\node[text width=.2cm](25) at (5.5,-2){$M_\alpha$};
\node[text width=.2cm](26) at (7.5,-2){$M_{\alpha +1}$};
\draw[thick] (1) -- (2);
\draw[thick] (2) -- (3);
\draw[thick] (3) -- (4);
\draw[thick,dashed] (4) -- (5);
\draw[thick] (5) -- (6);
\draw[thick] (6) -- (7);
\draw[thick] (7) -- (8);
\draw[thick] (7) -- (27);
\draw[thick] (8) -- (9);
\draw[thick] (8) -- (10);
\draw[thick,dashed] (10) -- (11);
\draw[thick,dashed] (1) -- (100);
\draw[thick] (2) -- (0,1.5);
\draw[thick, dashed] (0,1.5) -- (0,2.5);
\draw[thick] (2,1.5) -- (3);
\draw[thick, dashed] (2,1.5) -- (2,2.5);
\draw[thick] (5,1.5) -- (6);
\draw[thick, dashed] (5,1.5) -- (5,2.5);
\draw[thick] (7,1.5) -- (7);
\draw[thick, dashed] (7,1.5) -- (7, 2.5);
\draw[thick] (9,1.5) -- (8);
\draw[thick,dashed] (9,1.5) -- (9,2.5);
\draw[thick] (2) -- (20);
\draw[thick] (3) -- (21);
\draw[thick] (6) -- (22);
\end{tikzpicture}}
\end{center}
\caption{\footnotesize{A generic quiver with unitary/special unitary gauge nodes and (bi)fundamental matter. 
Black and yellow circular nodes denote unitary and special unitary gauge nodes respectively, black boxes 
denote fundamental hypers, and black lines connecting two gauge nodes denote bifundamental hypers. }}
\label{fig: USUgen}
\end{figure}
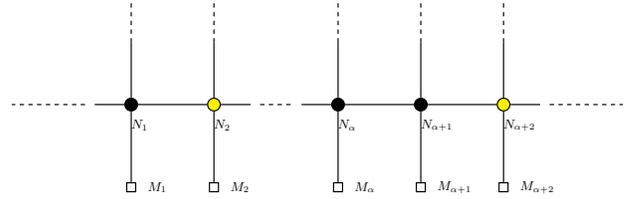

In this paper, we present a field theory construction of 3d mirrors for quiver gauge theories with unitary as well as 
special unitary gauge nodes and hypermultiplets in the fundamental/bifundamental representations, 
as shown in \figref{fig: USUgen}. Our algorithm does not rely on any String Theory realization of these theories, and may be applied 
to quiver gauge theories given by generic graphs, as long as the theories are good or ugly in the Gaiotto-Witten sense.  
The diagnosis of bad theories in this class of quivers and the construction of associated 3d mirrors will be discussed in a separate paper. \\

For a generic quiver of the form given in \Figref{fig: USUgen}, a unitary gauge node $U(N_\alpha)$ is \textit{balanced} if the 
balance parameter $e_\alpha = N^\alpha_{f/bf} - 2N_\alpha =0$, where $N^\alpha_{f/bf}$ is the total number of fundamental 
and bifundamental hypers associated with the node. In contrast, a special unitary node $SU(N_\alpha)$ is balanced if 
the balance parameter $e_\alpha = N^\alpha_{f/bf} - 2N_\alpha = -1$. A quiver with at least one balanced 
special unitary node has an emergent 0-form global symmetry on the CB \cite{Dey:2022pqr, Dey:2022abc, Dey:2023xhq} in the IR. For such a theory, the rank of 
the IR symmetry and that of the UV-manifest symmetry (realized as topological symmetries associated with the unitary gauge groups) 
are different, and the IR SCFT has mass deformations which are not visible in the UV Lagrangian. 

In \cite{Dey:2022abc, Dey:2023xhq}, such theories were shown to admit a sequence of IR dualities which 
are distinct from 3d mirror symmetry. Each of these IR dualities maps the CB and the HB of one theory to the respective CB and the HB 
of the dual theory in the deep IR. Therefore, the duality sequence leads to a set of quiver gauge theories which flow to the same IR SCFT and have 
isomorphic CBs and HBs in the deep IR. 
This phenomenon was referred to as \textit{IR N-ality}. For a quiver gauge theory $\CT$ of the form given in \figref{fig: USUgen} 
with one or more balanced special unitary nodes, it was also shown that there exists a preferred N-al theory $\CT_{\rm maximal}$ (generically not unique) for which the rank of the UV-manifest CB symmetry is maximized. If the quiver $\CT$ is given by a tree-graph (i.e. without any loops), the 
rank of this maximal UV-manifest CB symmetry is equal to the rank of the emergent CB symmetry. The theory $\CT_{\rm maximal}$ 
will play a crucial role in our construction of the 3d mirror of $\CT$.

\begin{figure}[htbp]
\begin{center}
\scalebox{0.5}{\begin{tikzpicture}
\node[] (100) at (-3,0) {};
\node[] (1) at (-1,0) {};
\node[unode] (2) at (0,0) {};
\node[text width=.2cm](31) at (0.1,-0.5){$N_1$};
\node[unode] (3) at (2,0) {};
\node[text width=.2cm](32) at (2.1,-0.5){$N_2$};
\node[] (4) at (3,0) {};
\node[] (5) at (4,0) {};
\node[unode] (6) at (5,0) {};
\node[text width=.2cm](33) at (5.1,-0.5){$N_{\alpha}$};
\node[unode] (7) at (7,0) {};
\node[text width=.2cm](34) at (7.1,-0.5){$N_{\alpha+1}$};
\node[text width=.2cm](35) at (5.5,0.3){$M_{\alpha\,\alpha+1}$};
\node[sunode] (8) at (9,0) {};
\node[text width=.2cm] (40) at (9.1,-0.5) {$N_{\alpha+2}$};
\node[fnode] (9) at (9,-2) {};
\node[text width=.2cm] (40) at (9.5,-2) {$M_{\alpha+2}$};
\node[] (10) at (10,0) {};
\node[] (11) at (12,0) {};
\node[afnode] (12) at (7,2) {};
\node[text width=.2cm](36) at (7.5,2){$F$};
\node[fnode] (20) at (0,-2) {};
\node[fnode] (21) at (2,-2) {};
\node[fnode] (22) at (5,-2) {};
\node[text width=.2cm](23) at (0.5,-2){$M_1$};
\node[text width=.2cm](24) at (2.5,-2){$M_2$};
\node[text width=.2cm](25) at (5.5,-2){$M_\alpha$};
\draw[thick] (1) -- (2);
\draw[line width=0.75mm, blue] (2) -- (3);
\node[text width=.2cm](50) at (0.5,0.3){$(\wt{Q}^1,\wt{Q}^2)$};
\node[text width=.2cm](51) at (1,-0.3){$P$};
\draw[thick] (3) -- (4);
\draw[thick,dashed] (4) -- (5);
\draw[thick] (5) -- (6);
\draw[line width=0.75mm] (6) -- (7);
\draw[thick] (7) -- (8);
\draw[thick,blue] (7) -- (12);
\draw[thick] (8) -- (9);
\draw[thick] (8) -- (10);
\draw[thick,dashed] (10) -- (11);
\draw[thick,dashed] (1) -- (100);
\draw[thick, blue] (2) -- (0,1.5);
\draw[thick, blue] (0,1.5) -- (5,1.5);
\draw[thick, blue] (2,1.5) -- (3);
\draw[thick, blue] (5,1.5) -- (6);
\draw[thick] (2) -- (20);
\draw[thick] (3) -- (21);
\draw[thick] (6) -- (22);
\node[text width=.2cm](15) at (0.25,1){$Q^1$};
\node[text width=.2cm](16) at (2.25,1){$Q^2$};
\node[text width=.2cm](17) at (5.25,1){$Q^\alpha$};
%\node[text width=.2cm](18) at (7.25,1){$Q^{\alpha+1}$};
\end{tikzpicture}}
\end{center}
\caption{\footnotesize{A generic quiver with unitary/special unitary gauge nodes, with (bi)fundamental and Abelian hypermultiplets. 
The charges $\{Q^i =\pm N_i\}$ encode whether the Abelian hypermultiplet transforms in the determinant or the anti-determinant  representation of 
a given unitary gauge node with $N_i$ being the rank of the unitary gauge node $i$.}}
\label{fig: AbHMgen}
\end{figure}
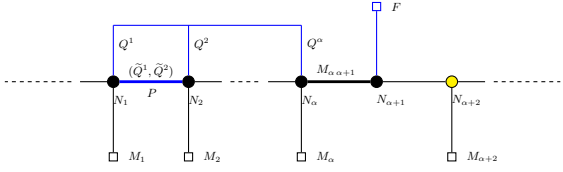

As discussed in \cite{Dey:2022abc, Dey:2023xhq}, the theory $\CT_{\rm maximal}$ (as well other theories N-al to $\CT$) is generically not of the type shown in \figref{fig: USUgen}. In addition to unitary/special unitary gauge nodes and fundamental/bifundamental matter, the theory $\CT_{\rm maximal}$ has hypermultiplets that transform in powers of the determinant and/or the anti-determinant representations of the unitary 
gauge nodes. We will collectively refer to these matter multiplets as \textit{Abelian hypermultiplets}. The quiver notation for a 
generic theory in this class is shown in \figref{fig: AbHMgen}. A blue square box with label $F$ represents $F$ 
hypermultiplets in the determinant representation of the unitary gauge node. A thin blue line connecting two or more 
unitary gauge nodes is an Abelian hypermultiplet associated with those gauge nodes, while a thick blue line with a label 
$P$ denotes a collection of $P$ Abelian hypermultiplets.\\

%The simplest example of a quiver gauge theory of the type shown in \figref{fig: AbHMgen} is a $U(N)$ quiver gauge theory 
%with $N_f$ fundamental hypers and $P$ Abelian hypers in the determinant representation -- we will denote them as 
%$\CT^N_{N_f,P}$. It can be shown \cite{} that these theories are good in the Gaiotto-Witten sense for $N_f \geq 2N-1$ 
%and any $P \geq 1$, and bad for $N_f < 2N-1$ for any $P$, while no ugly regime exists for $P \neq 0$. 

%\section{Finding $\CT_{\rm maximal}$ : Quiver Mutations and IR Duality Sequence}\label{CTMax}

\noindent \textit{Quiver Mutations and IR Duality Sequence.} 
Let us now briefly review the construction of the theory $\CT_{\rm maximal}$ given a quiver gauge theory $\CT$ with at least one balanced special unitary gauge node. For more details on the construction, we refer the reader to \cite{Dey:2022abc}. 

%\be \label{eq: Mutations I}
\begin{figure}[htbp]
\begin{tabular}{c}
\scalebox{0.6}{\begin{tikzpicture}
\node[] (1) at (1,0){};
\node[] (100) at (0,0){};
\node[unode] (2) at (2,0){};
\node[sunode] (3) at (4,0){};
\node[unode] (4) at (6,0){};
\node[unode] (51) at (2,2){};
\node[] (52) at (0,2){};
\node[] (53) at (1,2){};
\node[unode] (61) at (6,2){};
\node[] (62) at (7,2){};
\node[] (63) at (8,2){};
\node[] (5) at (7,0){};
\node[] (200) at (8,0){};
\node[cross, red] (6) at (4,0.5){};
%\node[fnode] (8) at (4,-2){};
\draw[-] (1) -- (2);
\draw[-] (2)-- (3);
\draw[-] (3) -- (4);
\draw[-] (4) --(5);
%\draw[-] (3) --(8);
\draw[-] (3) --(51);
\draw[-] (3) --(61);
\draw[-, dotted] (1) -- (100);
\draw[-,dotted] (5) -- (200);
\draw[-, dotted] (52) -- (53);
\draw[-] (51) -- (53);
\draw[-, dotted] (62) -- (63);
\draw[-] (61) -- (62);
\node[text width=.2cm](11) [below=0.1cm of 2]{$N_{\alpha_2}$};
\node[text width=.2cm](12) at (4.1, -0.5){$N_\alpha$};
\node[text width=.2cm](13) [below=0.1cm of 4]{$N_{\alpha_3}$};
%\node[text width=.2cm](14) [right=0.1cm of 8]{$M_\alpha$};
\node[text width=.2cm](15) [above=0.1cm of 51]{$N_{\alpha_1}$};
\node[text width=.2cm](16) [above=0.1cm of 61]{$N_{\alpha_4}$};
\node[text width=.2cm](20) at (-1,0){$(\CT)$};
\end{tikzpicture}} \\
 \scalebox{.6}{\begin{tikzpicture}
\draw[thick, ->] (0,0) -- (0,-1);
\node[text width=0.1cm](29) at (0. 3, -0.5) {$\CO_{I}$};
\node[](30) at (-1.5, -.5) {};
\end{tikzpicture}} \\
\scalebox{0.6}{\begin{tikzpicture}
\node[] (1) at (1,0){};
\node[] (100) at (0,0){};
\node[unode] (2) at (2,0){};
\node[unode] (3) at (4,0){};
\node[unode] (4) at (6,0){};
\node[unode] (51) at (2,2){};
\node[] (52) at (0,2){};
\node[] (53) at (1,2){};
\node[unode] (61) at (6,2){};
\node[] (62) at (7,2){};
\node[] (63) at (8,2){};
\node[] (5) at (7,0){};
\node[] (200) at (8,0){};
%\node[fnode] (8) at (4,-2){};
\draw[-] (1) -- (2);
\draw[-] (2)-- (3);
\draw[-] (3) -- (4);
\draw[-] (4) --(5);
%\draw[-] (3) --(8);
\draw[-] (3) --(51);
\draw[-] (3) --(61);
\draw[-, dotted] (1) -- (100);
\draw[-,dotted] (5) -- (200);
\draw[-, dotted] (52) -- (53);
\draw[-] (51) -- (53);
\draw[-, dotted] (62) -- (63);
\draw[-] (61) -- (62);
\draw[-, thick, blue] (2)--(2,1.5);
\draw[-, thick, blue] (3)--(4,1.5);
\draw[-, thick, blue] (4)--(6,1.5);
\draw[-, thick, blue] (2,1.5)--(4,1.5);
\draw[-, thick, blue] (4,1.5)--(6,1.5);
\draw[-, thick, blue] (4,1.5)--(51);
\draw[-, thick, blue] (4,1.5)--(61);
\node[text width=.2cm](11) [below=0.1cm of 2]{$N_{\alpha_2}$};
\node[text width=1.5cm](12) at (4.1, -0.5){$N_\alpha -1$};
\node[text width=.2cm](13) [below=0.1cm of 4]{$N_{\alpha_3}$};
%\node[text width=.2cm](14) [right=0.1cm of 8]{$M_\alpha$};
\node[text width=.2cm](15) [above=0.1cm of 51]{$N_{\alpha_1}$};
\node[text width=.2cm](16) [above=0.1cm of 61]{$N_{\alpha_4}$};
\node[text width=.2cm](20) at (-1, 0){$(\CT^\vee)$};
\end{tikzpicture}}\\
\qquad \\
\hline\\
\qquad\\
\scalebox{0.6}{\begin{tikzpicture}
\node[] (1) at (1,0){};
\node[] (100) at (0,0){};
\node[unode] (2) at (2,0){};
\node[unode] (3) at (4,0){};
\node[unode] (4) at (6,0){};
\node[unode] (51) at (2,2){};
\node[] (52) at (0,2){};
\node[] (53) at (1,2){};
\node[unode] (61) at (6,2){};
\node[] (62) at (7,2){};
\node[] (63) at (8,2){};
\node[] (5) at (7,0){};
\node[] (200) at (8,0){};
\node[cross, red] (6) at (4,0.5){};
%\node[fnode] (8) at (4,-2){};
\draw[-] (1) -- (2);
\draw[-] (2)-- (3);
\draw[-] (3) -- (4);
\draw[-] (4) --(5);
%\draw[-] (3) --(8);
\draw[-] (3) --(51);
\draw[-] (3) --(61);
\draw[-, dotted] (1) -- (100);
\draw[-,dotted] (5) -- (200);
\draw[-, dotted] (52) -- (53);
\draw[-] (51) -- (53);
\draw[-, dotted] (62) -- (63);
\draw[-] (61) -- (62);
\draw[dotted, thick, blue] (0,1.5)--(1,1.5);
\draw[-, thick, blue] (1,1.5)--(2,1.5);
\draw[-, thick, blue] (2)--(2,1.5);
\draw[-, thick, blue] (3)--(4,1.5);
\draw[-, thick, blue] (4)--(6,1.5);
\draw[-, thick, blue] (2,1.5)--(4,1.5);
\draw[-, thick, blue] (4,1.5)--(6,1.5);
\draw[-, thick, blue] (4,1.5)--(51);
\draw[-, thick, blue] (4,1.5)--(61);
\draw[-, thick, blue] (7,1.5)--(6,1.5);
\draw[dotted, thick, blue] (8,1.5)--(7,1.5);
\node[text width=.2cm](11) [below=0.1cm of 2]{$N_{\alpha_2}$};
\node[text width=.2cm](12) at (4.1, -0.5){$N_\alpha$};
\node[text width=.2cm](13) [below=0.1cm of 4]{$N_{\alpha_3}$};
%\node[text width=.2cm](14) [right=0.1cm of 8]{$M_\alpha$};
\node[text width=.2cm](15) [above=0.1cm of 51]{$N_{\alpha_1}$};
\node[text width=.2cm](16) [above=0.1cm of 61]{$N_{\alpha_4}$};
\node[text width=0.1 cm](10) at (4, 2){\footnotesize{$\vec Q$}};
\node[text width=.2cm](20) at (-1,0){$(\CT)$};
\end{tikzpicture}} \\
\scalebox{.6}{\begin{tikzpicture}
\draw[->] (0,0) -- (0,-1);
\node[text width=0.1cm](29) at (0.3, -0.5) {$\CO_{III}$};
\node[](30) at (-1.5, -.5) {};
\end{tikzpicture}}\\
 \scalebox{0.6}{\begin{tikzpicture}
\node[] (1) at (1,0){};
\node[] (100) at (0,0){};
\node[unode] (2) at (2,0){};
\node[unode] (3) at (4,0){};
\node[unode] (4) at (6,0){};
\node[unode] (9) at (4,3.5){};
\node[fnode] (10) at (2,3.5){};
\node[unode] (51) at (2,2){};
\node[] (52) at (0,2){};
\node[] (53) at (1,2){};
\node[unode] (61) at (6,2){};
\node[] (62) at (7,2){};
\node[] (63) at (8,2){};
\node[] (5) at (7,0){};
\node[] (200) at (8,0){};
%\node[fnode] (8) at (4,-2){};
\draw[-] (1) -- (2);
\draw[-] (2)-- (3);
\draw[-] (3) -- (4);
\draw[-] (4) --(5);
%\draw[-] (3) --(8);
\draw[-] (3) --(51);
\draw[-] (3) --(61);
\draw[-] (9) --(10);
\draw[-, dotted] (1) -- (100);
\draw[-,dotted] (5) -- (200);
\draw[-, dotted] (52) -- (53);
\draw[-] (51) -- (53);
\draw[-, dotted] (62) -- (63);
\draw[-] (61) -- (62);
\draw[dotted, thick, blue] (0,1.5)--(1,1.5);
\draw[-, thick, blue] (1,1.5)--(2,1.5);
\draw[-, thick, blue] (2)--(2,1.5);
\draw[-, thick, blue] (3)--(4,1.5);
\draw[-, thick, blue] (4)--(6,1.5);
\draw[-, thick, blue] (2,1.5)--(4,1.5);
\draw[-, thick, blue] (4,1.5)--(6,1.5);
\draw[-, thick, blue] (4,1.5)--(51);
\draw[-, thick, blue] (4,1.5)--(61);
\draw[-, thick, blue] (9)--(4,1.5);
\draw[-, thick, blue] (7,1.5)--(6,1.5);
\draw[dotted, thick, blue] (8,1.5)--(7,1.5);
\node[text width=.2cm](11) [below=0.1cm of 2]{$N_{\alpha_2}$};
\node[text width=1.5cm](12) at (4.1, -0.5){$N_\alpha -1$};
\node[text width=.2cm](13) [below=0.1cm of 4]{$N_{\alpha_3}$};
%\node[text width=.2cm](14) [right=0.1cm of 8]{$M_\alpha$};
\node[text width=.2cm](24) [right=0.1cm of 9]{$1$};
\node[text width=.2cm](25) [left=0.1cm of 10]{$1$};
\node[text width=.2cm](15) [above=0.1cm of 51]{$N_{\alpha_1}$};
\node[text width=.2cm](16) [above=0.1cm of 61]{$N_{\alpha_4}$};
\node[text width=2 cm](30) at (5.5, 2){\footnotesize{$(1,\vec Q')$}};
\node[text width=.2cm](20) at (-1, 0){$(\CT^\vee)$};
\end{tikzpicture}}
\end{tabular}
\caption{\footnotesize{The mutation $\CO_I$ and the mutation $\CO_{III}$ (for $P=1$).}}
\label{fig: Mutations}
\end{figure}
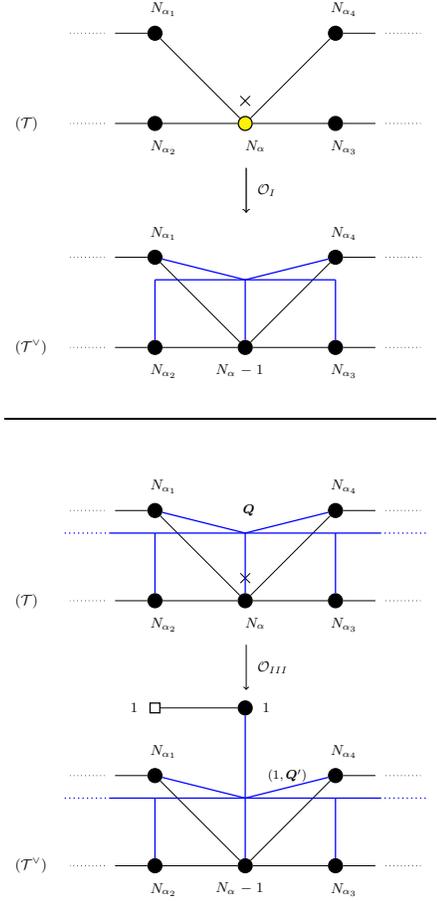
%\ee

The theory $\CT_{\rm maximal}$ can be obtained from the quiver $\CT$ by sequentially implementing a pair of quiver mutations, 
which we will refer to as mutation $I$ and mutation $III$ (and denote as $\CO_I$ and $\CO_{III}$ respectively in quiver diagrams). The former, 
shown in the top block of \Figref{fig: Mutations}, involves replacing a balanced $SU(N_\alpha)$ node by a $U(N_\alpha-1)$ node
and a single Abelian hypermultiplet with the following charge vector $\vec Q$:
\be \label{charge-1}
\vec Q=\Big(0,\ldots, N_{\alpha _1}, N_{\alpha _2}, -(N_\alpha -1), N_{\alpha _3}, N_{\alpha _4}, \ldots,0 \Big),
\ee
where $-(N_\alpha -1)$ is the charge under the $U(N_\alpha -1)$ gauge node, and $\{N_{\alpha_i}\}$ are the charges 
under those gauge nodes $\{U(N_{\alpha_i})\}$ which are connected to $U(N_\alpha -1)$ by bifundamental hypers. The 
hypermultiplet is not charged under any other gauge node in the quiver.

%The first mutation -- which we will refer to as mutation $I$ and denote as $\CO_I$ -- is shown in the top figure of \Figref{fig: Mutations}.
%The mutation involves replacing a balanced $SU$ node by a unitary node of the same rank and a single Abelian hypermultiplet. 
%The Abelian hyper in question has charge $-(N_\alpha -1)$ under the $U(N_\alpha -1)$ gauge node and is also charged under 
%the unitary gauge nodes connected to $U(N_\alpha -1)$ by bifundamental hypers. The charge vector associated with the Abelian 
%hyper has the generic form $\vec Q=(0,\ldots, N_{\alpha _1}, N_{\alpha _2}, -(N_\alpha -1), N_{\alpha _3}, N_{\alpha _4}, \ldots,0)$, 
%where $\{N_{\alpha_i}\}$ denote the ranks of the gauge nodes connected to $U(N_\alpha -1)$ by bifundamental hypers.

Mutation $III$, shown in the bottom block of \Figref{fig: Mutations}, involves replacing a $U(N_\alpha)$ gauge node, 
with balance parameter $e_\alpha = -1$ and $P \geq 1$ Abelian hypermultiplets, by a $U(N_\alpha-1)$ node and a $U(1)$ 
node where the latter node has a single fundamental hyper. In addition, the $P $ Abelian hypermultiplets get 
mapped by the mutation to $P $ Abelian hypermultiplets with different charges. Without loss of generality, 
we can choose the Abelian hypers in the quiver $\CT$ to have charge $N_\alpha$ under $U(N_\alpha)$ such that the charge vectors 
have the generic form 
\be \label{charge-3a}
\vec Q^l= \Big(Q^l_1, \ldots, Q^l_{\alpha_1}, Q^l_{\alpha_2}, N_\alpha, Q^l_{\alpha_3}, 
Q^l_{\alpha_4}, \ldots, Q^l_L \Big),
\ee
where $l=1,\ldots,P$, the total number of unitary nodes is $L$ and $\{Q^l_{\alpha_i}\}$ are the charges 
under those gauge nodes $\{U(N_{\alpha_i})\}$ which are connected to $U(N_\alpha)$ by bifundamental hypers. 
The charge vectors after the mutation are then given by:
\begin{align} \label{charge-3}
\vec Q'^l=\Big(& 1,Q^l_1, \ldots, Q^l_{\alpha_1} + N_{\alpha_1}, Q^l_{\alpha_2} + N_{\alpha_2},  -(N_\alpha-1), \nn \\
& Q^l_{\alpha_3} + N_{\alpha_3} , Q^l_{\alpha_4} + N_{\alpha_4}, \ldots, Q^l_L \Big),
\end{align}
where the first entry denotes the charge under the new $U(1)$ node and $-(N_\alpha-1)$ is the charge under the 
$U(N_\alpha -1)$ node. For the other nodes, only the charges associated with those connected with bifundamental 
hypers to $U(N_\alpha-1)$ get transformed under the mutation.

\begin{figure*}[htbp]
\scalebox{0.8}{\begin{tikzpicture}
  \node (D00) at (0,0) {$\wt{\CT}_{\rm good}$};
  \node (D10) at (4,0) {$\wt{\CT}_{\rm unitary}$};
  \node (D20) at (8,0) {$\wt{\CT}_{\rm maximal}$};
  \node (D01) at (0,3) {${\CT}_{\rm good}$};
  \node (D11) at (4,3) {${\CT}_{\rm unitary}$};
  \node (D21) at (8,3) {${\CT}_{\rm maximal}$};
  \node (D31) at (12,3) {$\CT$};
 \draw[->] (D00) -- (D10) node [midway, above] {\footnotesize Ab. $S$-type Op.};
 \draw[->] (D10) -- (D20) node [midway, above] {\footnotesize Ab. gauging};
   \draw[->] (D01) -- (D11) node [midway, above] {\footnotesize Ab. $S$-type Op.};
   \draw[->] (D11) -- (D21) node [midway, above] {\footnotesize Ab. gauging};
    \draw[->] (D31) -- (D21) node [midway, above] {\footnotesize IR dualities};
    \draw[->] (D00) -- (D01) node [midway, right=+3pt] {\footnotesize 3d mirror};
    \draw[->] (D01) -- (D00);
    \draw[->] (D10) -- (D11) node [midway, right=+3pt] {\footnotesize 3d mirror};
    \draw[->] (D11) -- (D10);
    \draw[->] (D20) -- (D21) node [midway, right=+3pt] {\footnotesize 3d mirror};
    \draw[->] (D21) -- (D20);
    \draw[->] (D20.north east) -- (D31) node [midway, right=+5pt] {\footnotesize 3d mirror};
    \draw[->] (D31) -- (D20.north east);
    \end{tikzpicture}}
\caption{\footnotesize{Construction of the 3d mirror.} }
\label{fig: flowchartMS}
\end{figure*}
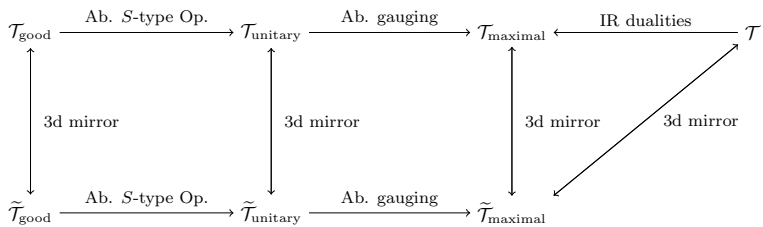

Given the quiver mutations $I$ and $III$, the theory $\CT_{\rm maximal}$ can be obtained in the following fashion. 
\begin{itemize}
\item We first implement mutation $I$ at each balanced $SU$ node in the quiver $\CT$. 

\item The mutation can render certain overbalanced $SU$ nodes balanced, and we implement mutation 
$I$ sequentially until we arrive at a quiver which has only overbalanced $SU$ nodes 
as well as unitary nodes with different balance parameters ($\geq -1$) plus a number of Abelian hypers.

\item In the next step, we implement mutation $III$ at the unitary gauge nodes 
that admit it. These operations will generically alter the balance parameters of both unitary and special unitary nodes in the quiver, 
thereby creating new nodes which admit mutation $III$ or mutation $I$. 

\item We implement these mutations sequentially until we arrive at a quiver where none of the gauge nodes admits 
either mutation $I$ or mutation $III$. This is where the duality sequence terminates and the resultant quiver is a candidate for $\CT_{\rm maximal}$.
\end{itemize}

The theory $\CT_{\rm maximal}$ obtained in this fashion is a quiver of generic type given in \Figref{fig: AbHMgen}, with the property 
that every $SU$ node is overbalanced, while every non-Abelian unitary node is either balanced or overbalanced. The quiver may 
contain certain $U(1)$ nodes which are attached to Abelian hypers and have balance -1, as these cannot be reduced any further. \\

%Note that the quiver operations $\CO_I$ and $\CO_{III}$ increase the number of $\fru(1)$ topological symmetries by 1, $\CO_{I'}$ decreases it by 1, 
%and $\CO_{II}$ keeps it invariant. This is why one can ignore $\CO_{I'}$ and $\CO_{II}$ if one is interested in finding a single 
%candidate for $\CT_{\rm maximal}$. However, the complete duality sequence must include these mutations as well. In particular, 
%there may be multiple candidates for $\CT_{\rm maximal}$ which are related by $\CO_{II}$. 
%

%\section{The prescription}\label{recipe}

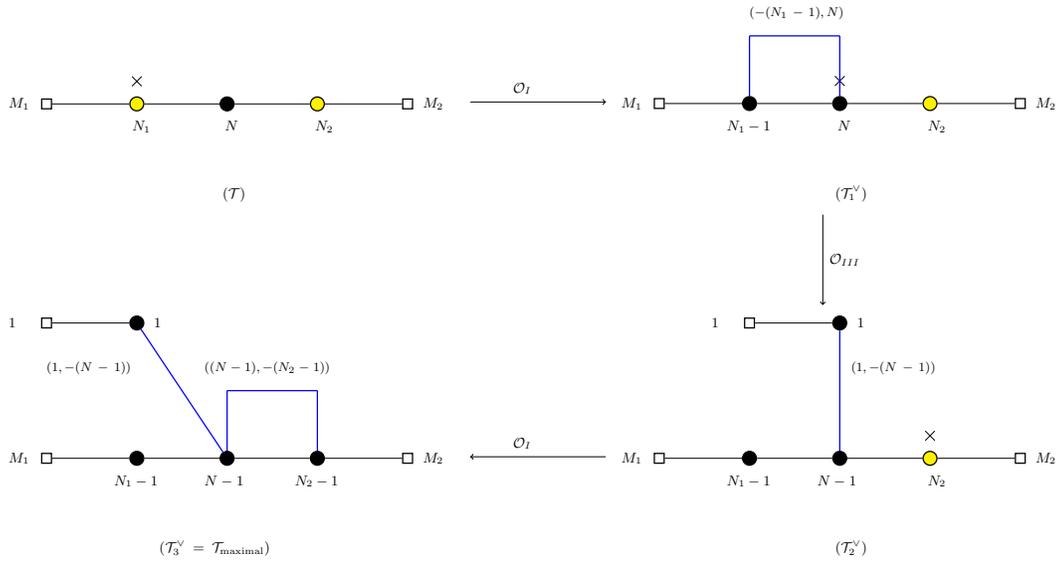
\begin{figure*}[htbp]
\begin{tabular}{ccc}
\scalebox{0.6}{\begin{tikzpicture}
\node[fnode] (1) at (0,0){};
\node[sunode] (2) at (2,0){};
\node[unode] (3) at (4,0){};
\node[sunode] (4) at (6,0){};
\node[fnode] (5) at (8,0){};
\node[cross, red] (6) at (2,0.5){};
\draw[-] (1) -- (2);
\draw[-] (2)-- (3);
\draw[-] (3) -- (4);
\draw[-] (4) --(5);
%\node[text width=0.5cm](12) [below=0.2cm of 2]{$\eta_1$};
%\node[text width=0.5cm](13) [below=0.2cm of 4]{$\eta_3$};
\node[text width=.1cm](10) [left=0.5 cm of 1]{$M_1$};
\node[text width=.2cm](11) [below=0.1cm of 2]{$N_1$};
\node[text width=.1cm](12) [below=0.1cm of 3]{$N$};
\node[text width=.1cm](13) [below=0.1cm of 4]{$N_2$};
\node[text width=.1cm](14) [right=0.1cm of 5]{$M_2$};
\node[text width=.2cm](20) at (4,-2){$(\CT)$};
\end{tikzpicture}}
& \scalebox{.6}{\begin{tikzpicture}
\node[] at (3.5,0){};
\draw[->] (4,0) -- (7, 0);
\node[text width=0.1cm](29) at (5, 0.3) {$\CO_{I}$};
\node[](30) at (5, -2.2) {};
\end{tikzpicture}}
&\scalebox{0.6}{\begin{tikzpicture}
\node[fnode] (1) at (0,0){};
\node[unode] (2) at (2,0){};
\node[unode] (3) at (4,0){};
\node[sunode] (4) at (6,0){};
\node[fnode] (5) at (8,0){};
\node[cross, red] (6) at (4,0.5){};
\draw[-] (1) -- (2);
\draw[-] (2)-- (3);
\draw[-] (3) -- (4);
\draw[-] (4) --(5);
\draw[-, thick, blue] (2)--(2,1.5);
\draw[-, thick, blue] (3)--(4,1.5);
%\draw[-, thick, blue] (4)--(6,1.5);
\draw[-, thick, blue] (2,1.5)--(4,1.5);
%\draw[-, thick, blue] (4,1.5)--(6,1.5);
\node[text width=3 cm](10) at (3.5, 2){\footnotesize{$(-(N_1-1), N)$}};
\node[text width=.1cm](10) [left=0.5 cm of 1]{$M_1$};
\node[text width= 1cm](11) [below=0.1cm of 2]{$N_1-1$};
\node[text width= 0.1cm](12) [below=0.1cm of 3]{$N$};
\node[text width=.1cm](13) [below=0.1cm of 4]{$N_2$};
\node[text width=.1cm](14) [right=0.1cm of 5]{$M_2$};
\node[text width=.2cm](20) at (4,-2){$(\CT^\vee_1)$};
\end{tikzpicture}}\\
\qquad 
& \qquad
& \scalebox{.6}{\begin{tikzpicture}
\draw[->] (4,-1) -- (4, -3);
\node[text width=0.1cm](29) at (4.2, -2) {$\CO_{III}$};
%\node[](30) at (1, -1.1) {};
\end{tikzpicture}}\\
\scalebox{0.6}{\begin{tikzpicture}
\node[fnode] (1) at (0,0){};
\node[unode] (2) at (2,0){};
\node[unode] (3) at (4,0){};
\node[unode] (4) at (6,0){};
\node[fnode] (5) at (8,0){};
\node[unode] (6) at (2,3){};
\node[fnode] (7) at (0,3){};
\draw[-] (1) -- (2);
\draw[-] (2)-- (3);
\draw[-] (3) -- (4);
\draw[-] (4) --(5);
\draw[-] (6) --(7);
\draw[-, thick, blue] (3)--(6);
\draw[-, thick, blue] (3)--(4,1.5);
\draw[-, thick, blue] (4)--(6,1.5);
%\draw[-, thick, blue] (4)--(6,1.5);
\draw[-, thick, blue] (4,1.5)--(6,1.5);
\node[text width=3 cm](40) at (1.5, 2){\footnotesize{$(1, -(N-1))$}};
\node[text width=3 cm](41) at (5, 2){\footnotesize{$((N-1), -(N_2-1))$}};
\node[text width=.1cm](20) [left=0.5 cm of 1]{$M_1$};
\node[text width=1 cm](21) [below=0.1cm of 2]{$N_1-1$};
\node[text width=1 cm](22) [below=0.1cm of 3]{$N-1$};
\node[text width=1 cm](23) [below=0.1cm of 4]{$N_2-1$};
\node[text width=.1cm](24) [right=0.1cm of 5]{$M_2$};
\node[text width=.1cm](25) [right=0.1cm of 6]{1};
\node[text width=.1cm](26) [left=0.5 cm of 7]{1};
\node[text width= 3 cm](20) at (4,-2){$(\CT^\vee_3=\CT_{\rm maximal})$};
\end{tikzpicture}}
& \scalebox{.6}{\begin{tikzpicture}
\node[] at (3.5,0){};
\draw[->] (7,0) -- (4, 0);
\node[text width=0.1cm](29) at (5, 0.3) {$\CO_{I}$};
\node[](30) at (5, -2.2) {};
\end{tikzpicture}}
&\scalebox{0.6}{\begin{tikzpicture}
\node[fnode] (1) at (0,0){};
\node[unode] (2) at (2,0){};
\node[unode] (3) at (4,0){};
\node[sunode] (4) at (6,0){};
\node[fnode] (5) at (8,0){};
\node[unode] (6) at (4,3){};
\node[fnode] (7) at (2,3){};
\node[cross, red] (10) at (6,0.5){};
\draw[-] (1) -- (2);
\draw[-] (2)-- (3);
\draw[-] (3) -- (4);
\draw[-] (4) --(5);
\draw[-] (6) --(7);
\draw[-, thick, blue] (3)--(6);
\node[text width=3 cm](40) at (5.75, 2){\footnotesize{$(1, -(N-1))$}};
\node[text width=.1cm](20) [left=0.5 cm of 1]{$M_1$};
\node[text width=1 cm](21) [below=0.1cm of 2]{$N_1-1$};
\node[text width=1 cm](22) [below=0.1cm of 3]{$N-1$};
\node[text width=0.1 cm](23) [below=0.1cm of 4]{$N_2$};
\node[text width=.1cm](24) [right=0.1cm of 5]{$M_2$};
\node[text width=.1cm](25) [right=0.1cm of 6]{1};
\node[text width=.1cm](26) [left=0.5 cm of 7]{1};
\node[text width=.2cm](20) at (4,-2){$(\CT^\vee_2)$};
\end{tikzpicture}}
\end{tabular}
\caption{\footnotesize{Derivation of $\CT_{\rm maximal}$ for a linear quiver $\CT$.} }
\label{IRdual-Ex1-main}
\end{figure*}

\noindent \textit{Prescription for the 3d mirror.} We can now write down the general recipe for the construction of the 3d mirror of a quiver gauge theory $\CT$ in the class of theories given in \figref{fig: USUgen}. The construction involves the following steps:

\begin{enumerate}

\item  Given a quiver $\CT$ containing at least one balanced special unitary gauge node, we obtain the theory $\CT_{\rm maximal}$ 
following the procedure outlined above. If $\CT$ has any ugly unitary node (i.e. balance parameter $e=-1$), we first apply the Gaiotto-Witten duality \cite{Gaiotto:2008ak} to write $\CT$ in terms of a good quiver and a set of free twisted hypermultiplets, and then implement the quiver mutations on the good quiver. Also, if $\CT$ has no balanced special unitary node, we simply have $\CT = \CT_{\rm maximal}$. 

\item Given $\CT_{\rm maximal}$, we replace all the overbalanced special unitary nodes by unitary nodes of the same label, i.e. we replace $SU(N)$ by $U(N)$. We will refer to this theory as $\CT_{\rm unitary}$. $\CT_{\rm unitary}$ can be mapped back to  $\CT_{\rm maximal}$ by gauging the topological $\fru(1)$ symmetries of the replaced unitary gauge nodes. Note that $\CT_{\rm unitary}$ consists of unitary gauge nodes only, while the matter sector includes fundamental/bifundamental hypers as well as Abelian hypers. If $\CT_{\rm maximal}$ has no overbalanced special unitary node, we have $ \CT_{\rm maximal} = \CT_{\rm unitary}$. 

\item In the next step, we strip off from $\CT_{\rm unitary}$ the Abelian hypers and the Abelian quiver tails introduced by the quiver mutations 
$I$ and $III$. The resultant theory is a standard quiver gauge theory with unitary gauge nodes and matter in the fundamental/bifundamental 
representation, and may consist of a number of possibly ugly $U(1)$ nodes (coming from $U(1)$ nodes with balance -1 attached to Abelian hypers 
in $\CT_{\rm maximal}$). Using the Gaiotto-Witten duality to remove these ugly nodes, we obtain a theory where the gauge nodes are all either balanced or overbalanced. For tree graphs this balance condition ensures that the theory is good, and we will refer to it as $\CT_{\rm good}$. 
For a generic graph, one has to check if this theory is indeed good before proceeding further. 

The theory $\CT_{\rm unitary}$ can be obtained from $\CT_{\rm good}$ by attaching to the gauge nodes of $\CT_{\rm good}$ 
one or more Abelian hypers which are in turn connected to Abelian quiver tails. This can be performed by introducing a decoupled 
Abelian quiver tail $\CT_{\rm decoupled}$ with the theory $\CT_{\rm good}$ and gauging a certain linear combination of the topological 
symmetries of gauge nodes in $\CT_{\rm decoupled}$ and $\CT_{\rm good}$. This procedure can be thought of as mild generalization of 
the $S$-operation \cite{Witten:2003ya} on $\CT_{\rm good}$. 

\item We then find the 3d mirror of $\CT_{\rm good}$, which we will refer to as $\wt{\CT}_{\rm good}$. 
This theory might be difficult to find, but the problem greatly simplifies if $\CT_{\rm good}$ is an (affine) $A$ or $D$-type quiver 
or a quiver which is related to quivers in these classes by a set of Abelian gauging operations. In a generic situation, one can try 
to engineer $\wt{\CT}_{\rm good}$ using the technology of non-Abelian $S$-type operations \cite{Dey:2020hfe}. 

\item Given the mirror pair $\CT_{\rm good}$ and $\wt{\CT}_{\rm good}$, we perform a set of QFT operations on $\CT_{\rm good}$ as 
described above which takes us back to $\CT_{\rm unitary}$. On the dual side, these operations act on $\wt{\CT}_{\rm good}$ to produce a theory which is 3d mirror of $\CT_{\rm unitary}$ -- we will refer to this theory as $\wt{\CT}_{\rm unitary}$.

\item In the final step, we recall that $\CT_{\rm maximal}$ can be obtained from $\CT_{\rm unitary}$ by gauging certain topological $\fru(1)$ symmetries in the latter quiver. On the dual side, this amounts to gauging certain Abelian subgroups of the HB global symmetry of $\wt{\CT}_{\rm unitary}$. The resultant theory is the 3d mirror of $\CT_{\rm maximal}$ -- we will refer to this as $\wt{\CT}_{\rm maximal}$. Since $\CT_{\rm maximal}$ is related to $\CT$ by a sequence of IR dualities, the quiver $\wt{\CT}_{\rm maximal}$ is the 3d mirror of $\CT$ (up to the free hypermultiplets if any in Step 1).

\end{enumerate}

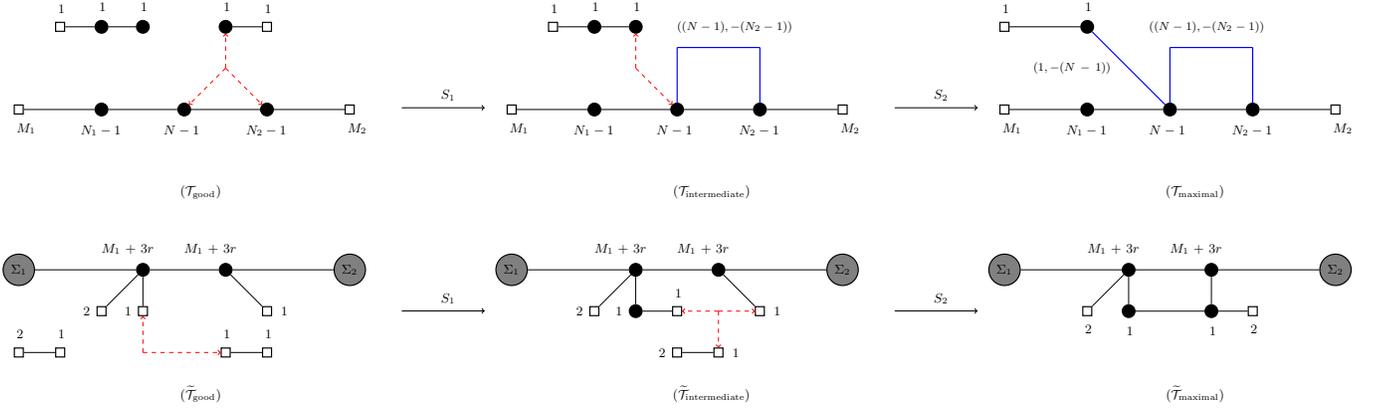
\begin{figure*}[t]
\begin{tabular}{ccccc}
\scalebox{0.55}{\begin{tikzpicture}
\node[fnode] (1) at (0,0){};
\node[unode] (2) at (2,0){};
\node[unode] (3) at (4,0){};
\node[unode] (4) at (6,0){};
\node[fnode] (5) at (8,0){};
\node[unode](101) at (5, 2){};
\node[fnode](102) at (6, 2){};
\node[unode](103) at (3, 2){};
\node[unode](104) at (2, 2){};
\node[fnode](105) at (1, 2){};
\draw[-] (1) -- (2);
\draw[-] (2)-- (3);
\draw[-] (3) -- (4);
\draw[-] (4) --(5);
\draw[-] (101) --(102);
\draw[-] (103) --(104);
\draw[-] (104) --(105);
\draw[->, red, dashed] (5,1) -- (101);
\draw[->, red, dashed] (5,1) -- (3);
\draw[->, red, dashed] (5,1) -- (4);
\node[text width=.1cm](20) [below=0.1 cm of 1]{$M_1$};
\node[text width=1 cm](21) [below=0.1cm of 2]{$N_1-1$};
\node[text width=1 cm](22) [below=0.1cm of 3]{$N-1$};
\node[text width=1 cm](23) [below=0.1cm of 4]{$N_2-1$};
\node[text width=.1cm](24) [below=0.1cm of 5]{$M_2$};
\node[text width=.1cm](25) [above=0.1cm of 101]{1};
\node[text width=.1cm](26) [above=0.1cm of 102]{1};
\node[text width=.1cm](27) [above=0.1cm of 103]{1};
\node[text width=.1cm](28) [above=0.1cm of 104]{1};
\node[text width=.1cm](29) [above=0.1cm of 105]{1};
\node[text width=.2cm](20) at (4,-2){$(\CT_{\rm good})$};
\end{tikzpicture}}
&\scalebox{.55}{\begin{tikzpicture}
\node[] at (3.5,0){};
\draw[->] (4,0) -- (6, 0);
\node[text width=0.1cm](29) at (5, 0.3) {$S_1$};
\node[](30) at (5, -2.2) {};
\end{tikzpicture}}
&\scalebox{0.55}{\begin{tikzpicture}
\node[fnode] (1) at (0,0){};
\node[unode] (2) at (2,0){};
\node[unode] (3) at (4,0){};
\node[unode] (4) at (6,0){};
\node[fnode] (5) at (8,0){};
\node[unode](101) at (3, 2){};
\node[unode](102) at (2, 2){};
\node[fnode](103) at (1, 2){};
\draw[-] (1) -- (2);
\draw[-] (2)-- (3);
\draw[-] (3) -- (4);
\draw[-] (4) --(5);
\draw[-] (101) --(102);
\draw[-] (102) --(103);
\draw[-, thick, blue] (3)--(4,1.5);
\draw[-, thick, blue] (4)--(6,1.5);
\draw[-, thick, blue] (4,1.5)--(6,1.5);
\draw[->, red, dashed] (3,1) -- (101);
\draw[->, red, dashed] (3,1) -- (3);
\node[text width=3 cm](41) at (5.5, 2){\footnotesize{$((N-1), -(N_2-1))$}};
\node[text width=.1cm](20) [below=0.1 cm of 1]{$M_1$};
\node[text width=1 cm](21) [below=0.1cm of 2]{$N_1-1$};
\node[text width=1 cm](22) [below=0.1cm of 3]{$N-1$};
\node[text width=1 cm](23) [below=0.1cm of 4]{$N_2-1$};
\node[text width=.1cm](24) [below=0.1cm of 5]{$M_2$};
\node[text width=.1cm](25) [above=0.1cm of 101]{1};
\node[text width=.1cm](26) [above=0.1cm of 102]{1};
\node[text width=.1cm](27) [above=0.1cm of 103]{1};
\node[text width=.2cm](20) at (4,-2){$(\CT_{\rm intermediate})$};
\end{tikzpicture}}
&\scalebox{.55}{\begin{tikzpicture}
\node[] at (3.5,0){};
\draw[->] (4,0) -- (6, 0);
\node[text width=0.1cm](29) at (5, 0.3) {$S_2$};
\node[](30) at (5, -2.2) {};
\end{tikzpicture}}
&\scalebox{0.55}{\begin{tikzpicture}
\node[fnode] (1) at (0,0){};
\node[unode] (2) at (2,0){};
\node[unode] (3) at (4,0){};
\node[unode] (4) at (6,0){};
\node[fnode] (5) at (8,0){};
\node[unode] (6) at (2,2){};
\node[fnode] (7) at (0,2){};
\draw[-] (1) -- (2);
\draw[-] (2)-- (3);
\draw[-] (3) -- (4);
\draw[-] (4) --(5);
\draw[-] (6) --(7);
\draw[-, thick, blue] (3)--(6);
\draw[-, thick, blue] (3)--(4,1.5);
\draw[-, thick, blue] (4)--(6,1.5);
%\draw[-, thick, blue] (4)--(6,1.5);
\draw[-, thick, blue] (4,1.5)--(6,1.5);
\node[text width=3 cm](40) at (2.2, 1){\footnotesize{$(1, -(N-1))$}};
\node[text width=3 cm](41) at (5, 2){\footnotesize{$((N-1), -(N_2-1))$}};
\node[text width=.1cm](20) [below=0.1 cm of 1]{$M_1$};
\node[text width=1 cm](21) [below=0.1cm of 2]{$N_1-1$};
\node[text width=1 cm](22) [below=0.1cm of 3]{$N-1$};
\node[text width=1 cm](23) [below=0.1cm of 4]{$N_2-1$};
\node[text width=.1cm](24) [below=0.1cm of 5]{$M_2$};
\node[text width=.1cm](25) [above=0.1cm of 6]{1};
\node[text width=.1cm](26) [above=0.1 cm of 7]{1};
\node[text width=.2cm](20) at (4,-2){$(\CT_{\rm maximal})$};
\end{tikzpicture}}\\
\qquad & \qquad & \qquad & \qquad & \qquad  \\
\scalebox{0.55}{\begin{tikzpicture}
\node[rnode] (1) at (0,0){$\Sigma_1$};
\node[unode] (2) at (3,0){};
\node[unode] (3) at (5,0){};
\node[rnode] (4) at (8,0){$\Sigma_2$};
\node[fnode] (5) at (3,-1){};
\node[fnode] (6) at (6,-1){};
\node[fnode] (7) at (2,-1){};
\node[fnode](101) at (5, -2){};
\node[fnode](102) at (6, -2){};
\node[fnode](103) at (1, -2){};
\node[fnode](104) at (0, -2){};
\draw[-] (1) -- (2);
\draw[-] (2)-- (3);
\draw[-] (3) -- (4);
\draw[-] (2) --(5);
\draw[-] (3) --(6);
\draw[-] (2) --(7);
\draw[-] (101) --(102);
\draw[-] (103) --(104);
\draw[->, red, dashed] (3,-2) -- (101);
\draw[->, red, dashed] (3,-2) -- (5);
\node[text width=2cm](20) [above=0.1 cm of 2]{$M_1+3r$};
\node[text width=2 cm](21) [above=0.1cm of 3]{$M_1+3r$};
\node[text width=0.1 cm](22) [left=0.1cm of 5]{1};
\node[text width=0.1 cm](23) [right=0.1cm of 6]{1};
\node[text width=0.1 cm](24) [left=0.1cm of 7]{2};
\node[text width=.1cm](25) [above=0.1cm of 101]{1};
\node[text width=.1cm](26) [above=0.1cm of 102]{1};
\node[text width=.1cm](27) [above=0.1cm of 103]{1};
\node[text width=.1cm](28) [above=0.1cm of 104]{2};
\node[text width=.2cm](20) at (4,-3){$(\wt{\CT}_{\rm good})$};
\end{tikzpicture}}
&\scalebox{.55}{\begin{tikzpicture}
\node[] at (3.5,0){};
\draw[->] (4,0) -- (6, 0);
\node[text width=0.1cm](29) at (5, 0.3) {$S_1$};
\node[](30) at (5, -2.2) {};
\end{tikzpicture}}
&\scalebox{0.55}{\begin{tikzpicture}
\node[rnode] (1) at (0,0){$\Sigma_1$};
\node[unode] (2) at (3,0){};
\node[unode] (3) at (5,0){};
\node[rnode] (4) at (8,0){$\Sigma_2$};
\node[unode] (5) at (3,-1){};
\node[fnode] (6) at (6,-1){};
\node[fnode] (7) at (2,-1){};
\node[fnode](101) at (4, -1){};
\node[fnode](103) at (5, -2){};
\node[fnode](104) at (4, -2){};
\draw[-] (1) -- (2);
\draw[-] (2)-- (3);
\draw[-] (3) -- (4);
\draw[-] (2) --(5);
\draw[-] (3) --(6);
\draw[-] (2) --(7);
\draw[-] (101) --(5);
\draw[-] (103) --(104);
\draw[->, red, dashed] (5,-1) -- (101);
\draw[->, red, dashed] (5,-1) -- (6);
\draw[->, red, dashed] (5,-1) -- (103);
\node[text width=2cm](20) [above=0.1 cm of 2]{$M_1+3r$};
\node[text width=2 cm](21) [above=0.1cm of 3]{$M_1+3r$};
\node[text width=0.1 cm](22) [left=0.1cm of 5]{1};
\node[text width=0.1 cm](23) [right=0.1cm of 6]{1};
\node[text width=0.1 cm](24) [left=0.1cm of 7]{2};
\node[text width=.1cm](25) [above=0.1cm of 101]{1};
\node[text width=.1cm](27) [right=0.1cm of 103]{1};
\node[text width=.1cm](28) [left=0.1cm of 104]{2};
\node[text width=.2cm](20) at (4,-3){$(\wt{\CT}_{\rm intermediate})$};
\end{tikzpicture}}
&\scalebox{.55}{\begin{tikzpicture}
\node[] at (3.5,0){};
\draw[->] (4,0) -- (6, 0);
\node[text width=0.1cm](29) at (5, 0.3) {$S_2$};
\node[](30) at (5, -2.2) {};
\end{tikzpicture}}
& \scalebox{0.55}{\begin{tikzpicture}
\node[rnode] (1) at (0,0){$\Sigma_1$};
\node[unode] (2) at (3,0){};
\node[unode] (3) at (5,0){};
\node[rnode] (4) at (8,0){$\Sigma_2$};
\node[unode] (5) at (3,-1){};
\node[fnode] (6) at (2,-1){};
\node[unode](7) at (5, -1){};
\node[fnode](8) at (6, -1){};
\draw[-] (1) -- (2);
\draw[-] (2)-- (3);
\draw[-] (3) -- (4);
\draw[-] (2) --(5);
\draw[-] (2) --(6);
\draw[-] (3) --(7);
\draw[-] (5) --(7);
\draw[-] (7) --(8);
\node[text width=2cm](20) [above=0.1 cm of 2]{$M_1+3r$};
\node[text width=2 cm](21) [above=0.1cm of 3]{$M_1+3r$};
\node[text width=0.1 cm](22) [below=0.1cm of 5]{1};
\node[text width=0.1 cm](23) [below=0.1cm of 6]{2};
\node[text width=0.1 cm](24) [below=0.1cm of 7]{1};
\node[text width=0.1 cm](25) [below=0.1cm of 8]{2};
\node[text width=.2cm](20) at (4,-3){$(\wt{\CT}_{\rm maximal})$};
\end{tikzpicture}}
\end{tabular}
\caption{Abelian quiver operations for constructing the 3d mirror of ${\CT}_{\rm maximal}$.}
\label{SOP-Ex1}
\end{figure*}

The above construction is schematically shown by a flow-chart in \figref{fig: flowchartMS}. Note that the sequence of IR dualities plays a crucial role 
in reducing the theory $\CT$ to a theory $\CT_{\rm good}$ which is given by a standard quiver with unitary gauge nodes. 
Assuming the 3d mirror of the latter is known, the 3d mirror of $\CT$ can be found by implementing a sequence of well-defined QFT operations 
involving Abelian gauging of global symmetries on the mirror pair $(\CT_{\rm good}, \wt{\CT}_{\rm good})$. In the rest of the paper, we illustrate the above recipe with a concrete example.\\

\noindent \textit{Illustrative Example : A Linear Quiver.} In this section, we will construct the 3d mirror of 
the following quiver gauge theory $\CT$ with two special unitary gauge nodes and a single unitary gauge node on a linear graph: 
\begin{center}
\scalebox{0.65}{\begin{tikzpicture}
\node[fnode] (1) {};
\node[sunode] (2) [right=.75cm  of 1]{};
\node[unode] (3) [right=.75cm of 2]{};
\node[sunode] (4) [right=0.75 cm of 3]{};
\node[fnode] (5) [right=0.75 cm of 4]{};
\draw[-] (1) -- (2);
\draw[-] (2)-- (3);
\draw[-] (3) -- (4);
\draw[-] (4) -- (5);
\node[text width=.1cm](10) [below=0.1 cm of 1]{$M_1$};
\node[text width=.2cm](11) [below=0.1cm of 2]{$N_1$};
\node[text width=.1cm](12) [below=0.1cm of 3]{$N$};
\node[text width=.1cm](13) [below=0.1cm of 4]{$N_2$};
\node[text width=.1cm](14) [below=0.1cm of 5]{$M_2$};
%\draw[-] (4) --(5);
\node[text width=.2cm](20) [left= 1 cm of 1]{$(\CT):$};
\end{tikzpicture}}
\end{center}

We will consider the case where the left $SU$ node and the central $U$ node are balanced while the right $SU$ node is overbalanced 
with balance parameter $e=0$. The integers $\{M_1, N_1,N,N_2,M_2\}$ therefore obey the following constraints : $M_1 + N=2N_1-1$, $N_1 + N_2 =2N$ and $M_2 + N =2N_2$, leading to a 2-parameter family of quiver gauge theories. For the sake of concreteness, we will assume that 
$M_1 < N_1 -1 < M_2$, with $N_1 -1 = M_1 + r$ for a positive integer $r$.

The first step is to determine the quiver $\CT_{\rm maximal}$. Starting from the quiver $\CT$, the duality sequence leading to $\CT_{\rm maximal}$ is given in \Figref{IRdual-Ex1-main} and involves the following steps:
\begin{enumerate}

\item We implement mutation $I$ on the balanced $SU(N_1)$ node of $\CT$ following \Figref{fig: Mutations} to obtain 
the quiver $\CT^\vee_1$. The charge of the Abelian hypermultiplet in $\CT^\vee_1$ can be read off from the equation \eref{charge-1}.

\item The $U(N)$ gauge node in $\CT^\vee_1$ is attached to an Abelian hypermultiplet and has balance parameter $e=-1$. Therefore, 
we can implement mutation $III$ at this node following \Figref{fig: Mutations} to obtain the quiver $\CT^\vee_2$. The charge of the Abelian hypermultiplet in $\CT^\vee_2$ can be read off from the equation \eref{charge-3}.

\item The $SU(N_2)$ gauge node of $\CT^\vee_2$ is balanced, which implies that we can implement another mutation $I$. This leads to the quiver 
$\CT^\vee_3$. 

\item The quiver $\CT^\vee_3$ has no special unitary nodes, while all the unitary nodes (whether connected to Abelian hypers or not) are 
either balanced or overbalanced. Therefore, the duality sequence terminates and we can identify $\CT^\vee_3=\CT_{\rm maximal}$. 

\end{enumerate}

The IR CB symmetry algebra of $\CT$ can be read off from the quiver 
$\CT_{\rm maximal}$. Noting that the $U(1)$ and the $U(N_1-1)$ gauge nodes are 
balanced while the other two are overbalanced, we have 
\be
\frg^{\rm IR}_{\rm C} (\CT_{\rm maximal})= \frg^{\rm IR}_{\rm C} (\CT) = \frsu(2) \oplus 
\frsu(2) \oplus \fru(1) \oplus \fru(1).
\ee

%\begin{center}
%\scalebox{0.6}{\begin{tikzpicture}
%\node[fnode] (1) at (0,0){};
%\node[unode] (2) at (2,0){};
%\node[unode] (3) at (4,0){};
%\node[unode] (4) at (6,0){};
%\node[fnode] (5) at (8,0){};
%\node[unode] (6) at (2,3){};
%\node[fnode] (7) at (0,3){};
%\draw[-] (1) -- (2);
%\draw[-] (2)-- (3);
%\draw[-] (3) -- (4);
%\draw[-] (4) --(5);
%\draw[-] (6) --(7);
%\draw[-, thick, blue] (3)--(6);
%\draw[-, thick, blue] (3)--(4,1.5);
%\draw[-, thick, blue] (4)--(6,1.5);
%%\draw[-, thick, blue] (4)--(6,1.5);
%\draw[-, thick, blue] (4,1.5)--(6,1.5);
%\node[text width=3 cm](40) at (1.5, 2){\footnotesize{$(1, -(N-1))$}};
%\node[text width=3 cm](41) at (5, 2){\footnotesize{$((N-1), -(N_2-1))$}};
%\node[text width=.1cm](20) [left=0.5 cm of 1]{$M_1$};
%\node[text width=1 cm](21) [below=0.1cm of 2]{$N_1-1$};
%\node[text width=1 cm](22) [below=0.1cm of 3]{$N-1$};
%\node[text width=1 cm](23) [below=0.1cm of 4]{$N_2-1$};
%\node[text width=.1cm](24) [right=0.1cm of 5]{$M_2$};
%\node[text width=.1cm](25) [right=0.1cm of 6]{1};
%\node[text width=.1cm](26) [left=0.5 cm of 7]{1};
%\node[text width=.2cm](20) at (-3,0){$(\CT_{\rm maximal})$};
%\end{tikzpicture}}
%\end{center}
The next step is to find the quiver $\CT_{\rm good} $. Since $\CT_{\rm maximal}$ does not have any special unitary node,
we have $\CT_{\rm unitary} = \CT_{\rm maximal}$. Stripping off the the Abelian hypers and attached Abelian 
quiver tails introduced by the quiver mutations from $\CT_{\rm maximal}$, we obtain the 
quiver $\CT_{\rm good} $: \\

\begin{center}
\scalebox{0.65}{\begin{tikzpicture}
\node[fnode] (1) at (0,0){};
\node[unode] (2) at (2,0){};
\node[unode] (3) at (4,0){};
\node[unode] (4) at (6,0){};
\node[fnode] (5) at (8,0){};
\draw[-] (1) -- (2);
\draw[-] (2)-- (3);
\draw[-] (3) -- (4);
\draw[-] (4) --(5);
\node[text width=.1cm](20) [left=0.5 cm of 1]{$M_1$};
\node[text width=1 cm](21) [below=0.1cm of 2]{$N_1-1$};
\node[text width=1 cm](22) [below=0.1cm of 3]{$N-1$};
\node[text width=1 cm](23) [below=0.1cm of 4]{$N_2-1$};
\node[text width=.1cm](24) [right=0.1cm of 5]{$M_2$};
\node[text width=.2cm](20) at (-3,0){$(\CT_{\rm good})$};
\end{tikzpicture}}
\end{center}

In the present case, the theory $\CT_{\rm good} $ is an $A$-type (linear) quiver which is manifestly good in the Gaiotto-Witten sense. 
This implies that the 3d mirror of $\CT_{\rm good}$ is also a good $A$-type quiver $\wt{\CT}_{\rm good}$, which can be 
found by the standard Hanany-Witten realization, and has the following form:

\begin{center}
\scalebox{0.65}{\begin{tikzpicture}
\node[rnode] (1) at (0,0){$\Sigma_1$};
\node[unode] (2) at (3,0){};
\node[unode] (3) at (5,0){};
\node[rnode] (4) at (8,0){$\Sigma_2$};
\node[fnode] (5) at (3,-2){};
\node[fnode] (6) at (5,-2){};
\draw[-] (1) -- (2);
\draw[-] (2)-- (3);
\draw[-] (3) -- (4);
\draw[-] (2) --(5);
\draw[-] (3) --(6);
\node[text width=2cm](20) [above=0.1 cm of 2]{$M_1+3r$};
\node[text width=2 cm](21) [above=0.1cm of 3]{$M_1+3r$};
\node[text width=0.1 cm](22) [left=0.1cm of 5]{3};
\node[text width=0.1 cm](23) [right=0.1cm of 6]{1};
\node[text width=.2cm](20) at (-3,0){$(\wt{\CT}_{\rm good})$};
\end{tikzpicture}}
\end{center}

The grey nodes, labelled $\Sigma_1$ and $\Sigma_2$, denote $A$-type quiver tails, and their explicit forms are given as follows: 

\begin{center}
\begin{tabular}{c}
\scalebox{0.55}{\begin{tikzpicture}
\node[rnode] (30) at (-2,0){$\Sigma_1$};
\node[text width=0.1 cm](100) at (-1,0){:};
\node[unode] (1) at (0,0){};
\node[unode] (2) at (2,0){};
\node[] (3) at (3,0){};
\node[] (4) at (5,0){};
\node[unode] (5) at (6,0){};
\node[unode] (6) at (8,0){};
\node[unode] (7) at (10,0){};
\node[] (8) at (11,0){};
\node[] (9) at (12,0){};
\node[unode] (10) at (13,0){};
\draw[-] (1) -- (2);
\draw[-] (2)-- (3);
\draw[-, dotted] (3) -- (4);
\draw[-] (4) -- (5);
\draw[-] (5) -- (6);
\draw[-] (6) -- (7);
\draw[-] (7) -- (8);
\draw[-, dotted] (8) -- (9);
\draw[-] (9) -- (10);
\node[text width=0.1 cm](20) [above=0.1 cm of 1]{$1$};
\node[text width= 0.1 cm](21) [above=0.1cm of 2]{$2$};
\node[text width=0.2 cm](22) [above=0.1cm of 5]{$M_1$};
\node[text width=1.5 cm](23) [above=0.1cm of 6]{$M_1+3$};
\node[text width=1.5 cm](24) [above=0.1cm of 7]{$M_1 + 6$};
\node[text width=2.5 cm](25) [above=0.1cm of 10]{$M_1 + 3(r-1)$};
\end{tikzpicture}}\\
\scalebox{0.55}{\begin{tikzpicture}
\node[rnode] (30) at (-4,0){$\Sigma_2$};
\node[text width=0.1 cm](100) at (-3,0){:};
\node[unode] (1) at (-1,0){};
\node[unode] (2) at (2,0){};
\node[] (3) at (3,0){};
\node[] (4) at (5,0){};
\node[unode] (5) at (7,0){};
\draw[-] (1) -- (2);
\draw[-] (2)-- (3);
\draw[-, dotted] (3) -- (4);
\draw[-] (4) -- (5);
\node[text width=2.5 cm](20) [above=0.1 cm of 1]{$M_1+3r -1$};
\node[text width= 2.5 cm](21) [above=0.1cm of 2]{$M_1+3r -2$};
\node[text width=0.1 cm](22) [above=0.1cm of 5]{$1$};
\end{tikzpicture}}
\end{tabular}
\end{center}

Note that the total number of gauge nodes in $(\wt{\CT}_{\rm good})$ is $M_1+M_2 -1$. One 
can readily check that global symmetry algebras across the duality agrees: $\frg^{\rm IR}_{\rm C} (\CT_{\rm good})= \frg^{\rm UV}_{\rm H} (\wt{\CT}_{\rm good})=\frsu(3) \oplus \fru(1)$, while $\frg^{\rm UV}_{\rm H} (\CT_{\rm good})=\frg^{\rm IR}_{\rm C} (\wt{\CT}_{\rm good}) = \frsu(M_1) \oplus \frsu(M_2) \oplus \fru(1)$, as expected.\\

The final step is to implement the Abelian $S$-type operations on $\CT_{\rm good}$ that leads to the quiver $\CT_{\rm maximal}$ -- 
the precise operations are shown in the top row of  \figref{SOP-Ex1}. The corresponding operations on the mirror side, which are 
shown in the bottom row, gives the 3d mirror of $\CT_{\rm maximal}$. We add to the quiver $\CT_{\rm good}$ a pair of 
decoupled Abelian quiver gauge theories : $$
\CT^{(1)}_{\rm decoupled} = \scalebox{0.7}{\begin{tikzpicture} 
\node[unode](101) at (5, 2){};
\node[fnode](102) at (6, 2){};
\draw[-] (101) --(102);
\node[text width=.1cm](25) [above=0.1cm of 101]{1};
\node[text width=.1cm](26) [above=0.1cm of 102]{1};
\end{tikzpicture}} 
\qquad 
\CT^{(2)}_{\rm decoupled} =\scalebox{0.7}{\begin{tikzpicture} 
\node[unode](103) at (3, 2){};
\node[unode](104) at (2, 2){};
\node[fnode](105) at (1, 2){}; 
\draw[-] (103) --(104);
\draw[-] (104) --(105);
\node[text width=.1cm](27) [above=0.1cm of 103]{1};
\node[text width=.1cm](28) [above=0.1cm of 104]{1};
\node[text width=.1cm](29) [above=0.1cm of 105]{1};
\end{tikzpicture}}$$ 
On the mirror side, we add the 3d mirrors of the aforementioned decoupled Abelian theories (a set of free hypermultiplets) to the 
theory $\wt{\CT}_{\rm good}$:
$$
\wt{\CT}^{(1)}_{\rm decoupled} = \scalebox{0.7}{\begin{tikzpicture}
\node[fnode](1) at (0, 0){};
\node[fnode](2) at (1, 0){};
\draw[-] (1) --(2);
\node[text width=.1cm](3) [above=0.1cm of 1]{1};
\node[text width=.1cm](4) [above=0.1cm of 2]{1};
\end{tikzpicture}} 
\qquad 
\wt{\CT}^{(2)}_{\rm decoupled} =\scalebox{0.7}{\begin{tikzpicture} 
\node[fnode](1) at (0, 0){};
\node[fnode](2) at (1, 0){};
\draw[-] (1) --(2);
\node[text width=.1cm](3) [above=0.1cm of 1]{2};
\node[text width=.1cm](4) [above=0.1cm of 2]{1};
\end{tikzpicture}}$$ 

In the first step, we gauge a linear combination of the $\fru(1)$ topological symmetries associated with a subset of
unitary gauge nodes in the quivers ${\CT}^{(1)}_{\rm decoupled}$ and ${\CT}_{\rm good}$. The gauge nodes in question
are marked by the red dashed arrows in the top left figure. This operation introduces an Abelian hyper charged under two nodes 
of the quiver $\CT_{\rm good}$, and gives the quiver $\CT_{\rm intermediate}$, as shown in the top middle figure. 
On the mirror side, this operation corresponds to identifying the Abelian flavor nodes marked by red dashed arrows in the bottom 
left figure and gauging the identified flavor node. This gives the quiver $\wt{\CT}_{\rm intermediate}$ as shown in the bottom middle figure 
-- the 3d mirror of $\CT_{\rm intermediate}$. 

In the next step, we perform an analogous operation on the quivers ${\CT}^{(1)}_{\rm decoupled}$ and $\CT_{\rm intermediate}$, 
as shown in the top middle figure, which gives $\CT_{\rm maximal}$.
The operation on the mirror side is shown in the bottom middle figure, and gives the quiver $\wt{\CT}_{\rm maximal}$ -- 
the 3d mirror of $\CT_{\rm maximal}$. One therefore ends up with the following mirror pair:
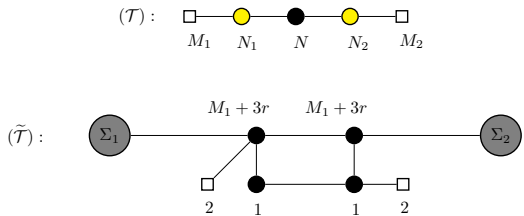
\begin{center}
\begin{tabular}{c}
\scalebox{0.65}{\begin{tikzpicture}
\node[fnode] (1) {};
\node[sunode] (2) [right=.75cm  of 1]{};
\node[unode] (3) [right=.75cm of 2]{};
\node[sunode] (4) [right=0.75 cm of 3]{};
\node[fnode] (5) [right=0.75 cm of 4]{};
\draw[-] (1) -- (2);
\draw[-] (2)-- (3);
\draw[-] (3) -- (4);
\draw[-] (4) -- (5);
\node[text width=.1cm](10) [below=0.1 cm of 1]{$M_1$};
\node[text width=.2cm](11) [below=0.1cm of 2]{$N_1$};
\node[text width=.1cm](12) [below=0.1cm of 3]{$N$};
\node[text width=.1cm](13) [below=0.1cm of 4]{$N_2$};
\node[text width=.1cm](14) [below=0.1cm of 5]{$M_2$};
%\draw[-] (4) --(5);
\node[text width=.2cm](20) [left= 1 cm of 1]{$(\CT):$};
\end{tikzpicture}}\\
\qquad\\
\scalebox{0.65}{\begin{tikzpicture}
\node[rnode] (1) at (0,0){$\Sigma_1$};
\node[unode] (2) at (3,0){};
\node[unode] (3) at (5,0){};
\node[rnode] (4) at (8,0){$\Sigma_2$};
\node[unode] (5) at (3,-1){};
\node[fnode] (6) at (2,-1){};
\node[unode](7) at (5, -1){};
\node[fnode](8) at (6, -1){};
\draw[-] (1) -- (2);
\draw[-] (2)-- (3);
\draw[-] (3) -- (4);
\draw[-] (2) --(5);
\draw[-] (2) --(6);
\draw[-] (3) --(7);
\draw[-] (5) --(7);
\draw[-] (7) --(8);
\node[text width=2cm](20) [above=0.1 cm of 2]{$M_1+3r$};
\node[text width=2 cm](21) [above=0.1cm of 3]{$M_1+3r$};
\node[text width=0.1 cm](22) [below=0.1cm of 5]{1};
\node[text width=0.1 cm](23) [below=0.1cm of 6]{2};
\node[text width=0.1 cm](24) [below=0.1cm of 7]{1};
\node[text width=0.1 cm](25) [below=0.1cm of 8]{2};
\node[text width=.2cm](20) at (-2, 0){$(\wt{\CT}):$};
\end{tikzpicture}}
\end{tabular}
\end{center}

Note that the emergent CB symmetry of the quiver $\CT$ is realized as a UV-manifest HB global symmetry in the 
theory $\wt{\CT}$, i.e.
\begin{align}
\frg^{\rm IR}_{\rm C} (\CT)= \frg^{\rm UV}_{\rm H}(\wt{\CT}) = \frsu(2) \oplus \frsu(2) \oplus \fru(1) \oplus \fru(1),
\end{align}
where the latter can be simply read off from the quiver $\wt{\CT}$. \\

%Let us now implement the Abelian field theory operations on the 3d mirror pair $(\CT_{\rm good}, \wt{\CT}_{\rm good})$. 
%The precise operations on the two theories are shown in the top row and the bottom row of \figref{SOP-Ex1}. In the first 
%step, we add a decoupled Abelian theory to $\CT_{\rm good}$ and gauge a combination of the $\fru(1)$ topological symmetries 
%associated to the unitary gauge nodes marked in the figure. This introduces an Abelian hyper in the quiver $\CT_{\rm good}$ and 
%leads to the quiver $\CT_{\rm intermediate}$. On the mirror side, we add the 3d mirror of the aforementioned decoupled Abelian theory 
%to $\wt{\CT}_{\rm good}$, and the gauging operation amounts to an identification and gauging of flavor nodes as shown. This leads to the 
%quiver $\wt{\CT}_{\rm intermediate}$ which is the 3d mirror of $\CT_{\rm intermediate}$. 

\noindent \textit{Conclusion and Outlook.} We have presented an explicit field theory algorithm for constructing the 3d mirror of 
a unitary-special unitary quiver gauge theory of generic shape with matter in the fundamental/bifundamental representation, 
where the quiver has at least a single balanced special unitary node and the theory is good or ugly in the Gaiotto-Witten sense. 
One can use a sequence of IR dualities to reduce a quiver $\CT$ in the aforementioned class 
to a quiver $\CT_{\rm maximal}$ which consists only of unitary gauge nodes but involves hypermultiplets in 
determinant/anti-determinant representation in addition to the fundamental/bifundamental matter. The quiver 
$\CT_{\rm maximal}$ can in turn be reduced to a unitary quiver gauge theory $\CT_{\rm good}$ with only fundamental/bifundamental 
hypermultiplets, by a sequence Abelian field theory operations. In this fashion, one reduces the problem of finding 
the 3d mirror of a unitary-special unitary quiver to the problem of finding the 3d mirror of a simpler unitary quiver. 
Given the 3d mirror of $\CT_{\rm good}$, we can retrace our steps and construct the 3d mirror of $\CT$.

We illustrate the construction by finding the mirror of a linear unitary-special unitary quiver gauge theory. 
The theory has an emergent CB symmetry in the IR, which appears in the 3d mirror as a UV-manifest 
Higgs branch symmetry. Generally speaking, if $\CT$ is a linear graph, then the quiver $\CT_{\rm good}$ 
is a linear quiver with unitary gauge nodes and therefore has a 3d mirror which is also a linear quiver 
with unitary gauge nodes. This makes the above construction particularly easy to implement.
However, examples of 3d mirrors for theories beyond linear quivers are reasonably straightforward 
to construct using this algorithm and some of them will appear in future papers.\\

It is well known that the circle reduction for a large class of Argyres-Douglas theories leads to unitary-special 
unitary quivers \cite{Closset:2020afy} of the type studied in this paper. The algorithm presented here 
therefore gives an alternative way to construct the 3d mirrors of circle-reduced Argyres-Douglas theories, 
without resorting to either the class $\CS$ technology or the magnetic quiver approach. 
This topic will be explored in an upcoming paper. \\

\noindent \textbf{Acknowledgments}  The author acknowledges the hospitality of the Simons Summer Workshop 
2023 during the completion of this work. The author is supported in part at the Johns Hopkins University by the 
NSF grant PHY-2112699.

\bibliographystyle{apsrev4-1}
\bibliography{cpn1-1}

\end{document}